\documentclass[submission,copyright,creativecommons]{eptcs}
\providecommand{\event}{PACO 2011} 
\usepackage{breakurl}             

\usepackage{subfigure}
\usepackage{amssymb,amsmath}
\usepackage{graphicx}
\usepackage{array}
\usepackage{tikz}

\hypersetup{
  colorlinks=true,
  urlcolor=green,
  citecolor=blue,
  linkcolor=violet,
  pdfpagemode=UseNone,
  pdfstartview=XYZ null null 1
}

\usetikzlibrary{arrows,shapes,automata,decorations,decorations.markings}

\tikzstyle{border} = [thin]

\tikzstyle{every state} =
	[circle,draw,border,inner sep=0pt,minimum size=5mm,initial text=]

\tikzstyle{transition} = [->,border,>=stealth']

\tikzstyle{point} = [thick,circle,inner sep=0.5pt,fill=black,draw=black,node distance=1cm]

\tikzstyle{channel}=[thick,>=stealth]
\tikzstyle{sync}=[channel,->]
\tikzstyle{lossysync}=[channel,->,dashed]
\tikzstyle{fifo}=[channel,->,
                  postaction=decorate,
                  decoration={markings,mark=at position 0.5 with
                  {\node[rectangle,draw=black,fill=white,transform shape]{$\quad$};}}]
\tikzstyle{syncdrain}=[channel,>-<]
\tikzstyle{asyncdrain}=[channel,>-<,
    postaction=decorate,
    decoration={markings,mark=at position 0.5 with
    {\draw[channel,-] (-2pt,-3pt) -- (-2pt,3pt) [transform shape]
                      (2pt,-3pt) -- (2pt,3pt) [transform shape];}}]

\newcommand{\noflowLeft}{%
\draw[border,solid,draw=black,fill=white](-2pt,-2.5pt) -- (-2pt,2.5pt) -- (2pt,0pt) -- cycle;}
\newcommand{\noflowRight}{%
\draw[border,solid,draw=black,fill=white](-2pt,0pt) -- (2pt,2.5pt) -- (2pt,-2.5pt) -- cycle;}

\tikzstyle{flow} = [border,thick] \tikzstyle{noflow}    = [border,dashed]
\tikzstyle{noflowLeft} = [noflow,postaction=decorate,
   decoration={markings,mark=at position 0.5 with \noflowLeft}]
\tikzstyle{noflowRight} = [noflow,postaction=decorate,
   decoration={markings,mark=at position 0.5 with \noflowRight}]
\tikzstyle{noflowLeftLeft} = [noflow,postaction=decorate,
   decoration={markings,mark=at position 0.25 with \noflowLeft,
                        mark=at position 0.75 with \noflowLeft}]
\tikzstyle{noflowRightRight} = [noflow,postaction=decorate,
   decoration={markings,mark=at position 0.25 with \noflowRight,
                        mark=at position 0.75 with \noflowRight}]
\tikzstyle{noflowLeftRight} = [noflow,postaction=decorate,
   decoration={markings,mark=at position 0.25 with \noflowLeft,
                        mark=at position 0.75 with \noflowRight}]

\newtheorem{definition}{Definition}[section]
\newtheorem{proposition}{Proposition}[section]

\newcommand{\mergerNode}{\tikz{
	    \node[inner sep=0,label=right:$C$] (D1) {};
	    \node[point,left of=D1,node distance=4mm] (C1) {};
	    \node[inner sep=0,label=left:$A$,above left of=C1,node
            distance=4mm] (B1) {};
	    \node[inner sep=0,label=left:$B$,below left of=C1,node
            distance=4mm] (A1) {};
	    \draw[sync] (C1) to (D1);
	    \draw[channel] (A1) to (C1);
	    \draw[channel] (B1) to (C1); }}

\newcommand{\replicatorNode}{\tikz{
	    \node[inner sep=0,label=left:$A$] (D1) {};
	    \node[point,right of=D1,node distance=4mm] (C1) {};
	    \node[inner sep=0,label=right:$B$,above right of=C1,node
          distance=4mm] (B1) {};
	    \node[inner sep=0,label=right:$C$,below right of=C1,node
          distance=4mm] (A1) {};
	    \draw[channel] (D1) to (C1);
	    \draw[sync] (C1) to (A1);
	    \draw[sync] (C1) to (B1); }}

\newcommand{\Reo}{\textsf{Reo}}
\renewcommand{\Reo}{Reo}
\newcommand{\mCRL}{\texttt{mCRL2}}

\newcommand{\A}{\ensuremath{\mathcal{A}}}

\newcommand{\N}{\ensuremath{\mathcal{N}}}
\newcommand{\M}{\ensuremath{\mathcal{M}}}

\newcommand{\calT}{\ensuremath{\mathcal{T}}}

\newcommand{\channel}[1]{\ensuremath{\mathsf{#1}}}

\newcommand{\AsyncDrain}{\channel{AsyncDrain}}
\newcommand{\FIFO}{\channel{FIFO}}
\newcommand{\Filter}{\channel{Filter}}

\newcommand{\LossySync}{\channel{LossySync}}

\newcommand{\JoinNode}{\channel{Join}}
\newcommand{\Sync}{\channel{Sync}}
\newcommand{\SyncDrain}{\channel{SyncDrain}}
\newcommand{\Transform}{\channel{Transform}}

\newcommand{\datatype}[1]{\ensuremath{\mathit{#1}}}
\newcommand{\Data}{\datatype{Data}}
\newcommand{\DataFIFO}{\datatype{DataFIFO}}
\newcommand{\Bool}{\datatype{Bool}}

\newcommand{\longtransition}[3]{\ensuremath{#1\overset{#2}{\xrightarrow{\hspace*{1.5cm}}}#3}}

\def\lparal{\mathbin{\setbox0=\hbox{$\|$}%
  \dimen0=\dp0 \advance\dimen0 -1.5pt \dp0=\dimen0%
  \underline{\kern-1.5pt\box0\kern1.5pt}}}

\newcommand{\transition}[3]{\ensuremath{#1\overset{#2}{\longrightarrow}#3}}

\newcommand{\notransition}[3]{\ensuremath{#1\overset{#2}{\nrightarrow}#3}}
\newcommand{\trace}[3]{\ensuremath{#1\overset{#2}{\Longrightarrow}#3}}

\newcommand{\mergeProc}{\mathbin{\parallel}}

\newcommand{\seq}{\ensuremath{\cdot}}
\newcommand{\choice}{\ensuremath{\;+\;}}

\newcommand{\struct}{\ensuremath{\mathbf{struct}}}

\newcommand{\mydot}{\mathbin{.}}

\newcommand{\ReplicatorNode}{\channel{ReplicatorNode}}

\newcommand{\shorten}[1]{}

\newcommand{\TT}[2]{{\ensuremath{\begin{array}{c} \{#1\} \\ #2 \end{array}}}}

\title{Input-output Conformance Testing for Channel-based Service Connectors}

\author{
  Natallia Kokash \qquad Farhad Arbab \qquad Behnaz Changizi
  \institute{
    CWI, P.O. Box 94079, 1090 GB Amsterdam, The Netherlands
  }
  \email{firstName.lastName@cwi.nl}
  \and Leonid Makhnist
  \institute{
    Brest State Technical University, Department of Higher Mathematics, Moskovskaya 267, 224017 Brest, Republic of Belarus
  }
}

\def\titlerunning{Input-output Conformance Testing for Channel-based Service Connectors}
\def\authorrunning{N. Kokash et al.}

\begin{document}
\maketitle

\begin{abstract}
Service-based systems are software systems composed of autonomous components or services provided by different vendors, deployed on remote machines and accessible through the web. One of the challenges of modern software engineering is to ensure that such a system behaves as intended by its designer. The Reo coordination language is an extensible notation for formal modeling and execution of service compositions. Services that have no prior knowledge about each other communicate through advanced channel connectors which guarantee that each participant, service or client, receives the right data at the right time. Each channel is a binary relation that imposes synchronization and data constraints on input and output messages. Furthermore, channels are composed together to realize arbitrarily complex behavioral protocols. During this process, a designer may introduce errors into the connector model or the code for their execution, and thus affect the behavior of a composed service. In this paper, we present an approach for model-based testing of coordination protocols designed in Reo. Our approach is based on the input-output conformance (ioco) testing theory and exploits the mapping of automata-based semantic models for Reo to equivalent process algebra specifications.
\end{abstract}


\section{Introduction}
\label{sect:Introduction}

Business process modeling is part of software development lifecycle which is primarily concerned with capturing the behavior of organizational business processes in a form that simplifies their analysis, fostering communication with various process stakeholders and helping to identify the requirements for the development of supporting software. Typically models are written using some (preferably standard) language or notation such as BPMN or UML diagrams. Once a process model has been constructed, it can be analyzed to uncover logical flaws in a process or optimize its functional or non-functional characteristics~\cite{Mor08,LVD09}.

While popular high-level modeling notations like BPMN or UML are suitable for fast prototyping and capturing system requirements, they are rather ambiguous and imprecise to be used for rigorous process analysis. Modeling languages should operate on the level of abstraction that allows designers to focus on the essence of the problem without being lost in technical details and at the same time provide sufficient precision and expressiveness to avoid ambiguities in the model or failure to describe certain important concepts. Multiple efforts on creating such modeling languages resulted into formalisms such as Petri nets and various process algebra-based languages often empowered with graphical syntax to simplify the process of unambiguous system description. These models are more difficult to use compared to high-level notations. However, their handicap of usability is compensated by automated validation and verification tools that provide powerful support for process analysis and quality assurance. Moreover, various model-based transformation tools have been developed for major notations to assist process designers with converting high-level process models into more rigorous ones.

Reo~\cite{Arb04:mscs} is an extensible model for coordination of software components or services wherein complex connectors are constructed out of simple primitives called channels. A channel is a binary relation that defines synchronization and data constraints on its input and output parameters. By composing basic channels, arbitrarily complex interaction protocols can be realized. Previous work shows that most of the behavioral patterns expressible in BPMN or UML notations can be modeled with Reo~\cite{AKS08}. We have also developed a set of tools for automated conversion of such models to Reo\footnote{\url{http://reo.project.cwi.nl/cgi-bin/trac.cgi/reo/wiki/Converters}}. Each Reo channel has a graphical representation and associated semantics. The most basic semantic model that currently exists for Reo relies on constraint automata~\cite{BSA+06}. Action constraint automata~\cite{KCA10} constitute a model that generalizes constraint automata by allowing more detailed observations on connector ports. When channels with timed, context-sensitive and probabilistic behavior are used to design a connector, more expressive models to represent the semantics of the connector are required~\cite{ABB+07,Bai05,BCS09}.

When using just a minimal set of channel types, it may happen that a substantial number of channels are required to construct a circuit with certain behavior. In general, it is not a trivial task to create a connector that implements a certain behavioral protocol. As any laborious process, connector implementation is error-prone and requires validation of the connector's behavior. There are several tools that can help to detect possible errors in Reo connectors. One of them is the animation engine~\cite{ECT}. This tool shows flash animated simulation of designed connectors and enables quick validation of connector designs. However, for complex connectors the number of possible traces is large and they are hard to analyze manually. Moreover, the current implementation of the animation engine is based on coloring semantics and cannot be used for reliable validation of data-dependent connectors.
A more efficient analysis of Reo models can be performed with the help of simulation and model-checking tools, both specifically developed for Reo~\cite{BBK+09,BI10,Kem09} and external~\cite{KKV11,KSA+08}. For example, simulation tools \emph{lpsxsim} and \emph{ocis} from {\mCRL}~\cite{Gro07:dagstuhl} and CADP~\cite{GML+07:cav} toolsets can be used to visualize execution traces of data-aware Reo networks followed by a user. Model checking tools \emph{pbes2bool} and \emph{evaluator} can be used to check the validity of connector properties expressed in the modal $\mu$-calculus formulae.

Both kinds of tools require substantial effort from the designer to analyze simulation traces or correctly express complex properties using the intricate $\mu$-calculus syntax. Yet another limitation of the aforementioned tools is their inability to analyze actual coordination code or protocol implementations. For example, in the context of the EU FP7 COMPAS project\footnote{\url{http://www.compas-ict.eu/}} we used Reo to design business process fragments and verify their conformance to various requirements extracted from compliance documents~\cite{STK+10}. These fragments are further implemented in BPEL and stored in a repository to enable their on-demand retrieval and reuse in service-based systems. While we can verify the correctness of Reo models in this scenario, we cannot judge the correctness of fragment implementations.

In this paper, we extend our previous work on verification of Reo with model-based testing facilities to automatically derive tests from connector specifications and execute them to test service coordination code or protocol implementations. We enable testing of connector designs given specifications of their expected behavior in the form of constraint automata extended with inputs and outputs. Test generation is based on the \emph{ioco}-testing theory which uses labelled transition systems (LTS) to represent system specifications, implementations and tests and defines a formal implementation relation called \emph{ioco} to show conformance between implementations and specifications. The encoding of automata-based behavioral semantics for Reo in process algebra {\mCRL} is exploited to obtain LTS models suitable for testing Reo. Together with previously developed tools for converting specifications in high-level process modeling notations such as BPMN and UML to Reo, graphical Reo networks can be used as a formal specification of business process models. In this case, Reo connectors are seen as formal specifications of processes and used to automatically derive tests to check the quality of process implementations. Since the \emph{ioco}-testing theory can be used to generate tests given specifications in any language with the LTS-based formal semantics, we can apply it to derive tests for any systems specified in Reo.

The remainder of this paper is organized as follows. In Section~\ref{sect:reo}, we explain the basics of Reo. In Section~\ref{sect:testingTheory}, we briefly summarize the basics of input-output conformance (\emph{ioco}) testing theory. In Section~\ref{sect:testingReo}, we explain how this theory can be used to test Reo. In Section~\ref{sect:toolSupport}, we illustrate the use of model-based testing tools to analyze Reo connectors. Finally, in Section~\ref{sect:conclusions}, we conclude the paper and outline our future work.

\section{The Reo Coordination Language}
\label{sect:reo}

\tikzstyle{every state}=[draw=black,text=black,minimum size=10pt,node
  distance=1.9cm,inner sep=1pt]
\tikzstyle{transition}=[->,>=stealth']
\newcommand{\T}[2]{{\ensuremath{\begin{array}{c} \{#1\} \\ #2 \end{array}}}}
\newcommand{\U}[2]{{\ensuremath{\{#1\} \; #2}}}

\begin{figure}[t]
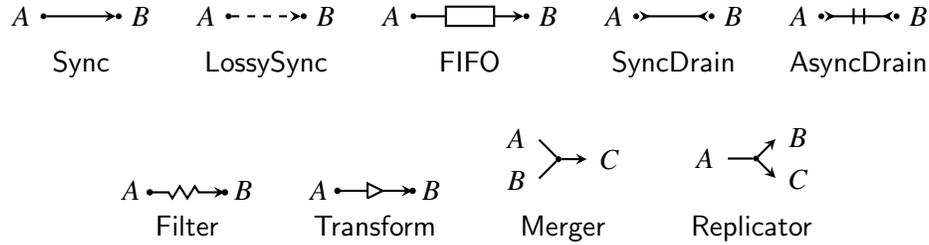

 \centering
 \scalebox{1.0}{
 \begin{tabular}{c}
     \begin{tabular}{c@{\quad}c@{\quad}c@{\quad}c@{\quad}c}
         \tikz{
    	   \node[point,label=left:$A$] (A) {};
    	   \node[point,right of=A,label=right:$B$] (B) {};
      	     \draw[sync] (A) -- (B); } &
         \tikz{
    	   \node[point,label=left:$A$] (A) {};
    	   \node[point,right of=A,label=right:$B$] (B) {};
      	     \draw[lossysync] (A) -- (B); } &
         \tikz{
            \node[point,label=left:$A$] (A) {};
            \node[point,right of=A,label=right:$B$, node distance=1.5cm] (B) {};
            \draw[fifo] (A) -- (B); } &
        \tikz{
            \node[point,label=left:$A$] (A) {};
            \node[point,right of=A,label=right:$B$] (B) {};
            \draw[syncdrain] (A) -- (B); } &
         \tikz{
            \node[point,label=left:$A$] (A) {};
            \node[point,right of=A,label=right:$B$] (B) {};
            \draw[asyncdrain] (A) -- (B); }
        \\
        \Sync & \LossySync & \FIFO & \SyncDrain & \AsyncDrain
      \end{tabular}
  \\
  \\
      \begin{tabular}{c@{\qquad}c@{\qquad}c@{\qquad}c}
          $A\;$\tikz{
            \filldraw[black] (0,0) circle (1pt) (1,0) circle (1pt);
            \draw[sync] (0,0) -- (0.25,0) -- (0.3, -0.08) -- (0.4, 0.08)
            -- (0.5, -0.08) -- (0.6,0.08) --(0.65, 0) -- (0.65,0) -- (1,0); }$\;B$ &
          $A\;$\tikz{
             \filldraw[black] (0,0) circle (1pt) (1,0) circle (1pt);
             \draw[channel,-] (0,0) -- (0.4,0);
             \draw[channel,-] (0.4, -0.1) -- (0.4, 0.1) -- (0.6, 0) --(0.4, -0.1);
             \draw[sync](0.6,0) -- (1,0); }$\;B$ &
          {\mergerNode} & {\replicatorNode}
        \\
        \Filter & \Transform & \channel{Merger} & \channel{Replicator}
      \end{tabular}
  \end{tabular}
  }
\caption{Graphical representation of basic Reo channels and nodes}
\label{fig:basicChannels}
\end{figure}

{\Reo} is a coordination language in which components and services are
coordinated exogenously by channel-based
connectors~\cite{Arb04:mscs}. Connectors are essentially graphs where
the edges are user-defined communication channels and the nodes
implement a fixed routing policy.  Channels in {\Reo} are entities
that have exactly two ends, also referred to as ports, which can be
either source or sink ends. Source ends accept data into, and sink
ends dispense data out of their respective channels. Although channels can be
defined by users, a set of basic Reo channels (see
Figure~\ref{fig:basicChannels}) with predefined behavior suffices to
implement rather complex coordination protocols. Among these channels
are (i) the {\Sync} channel, which is a directed channel that accepts
a data item through its source end if it can instantly dispense it
through its sink end; (ii) the {\LossySync} channel, which always
accepts a data item through its source end and tries to instantly
dispense it through its sink end. If this is not possible, the data
item is lost; (iii) the {\SyncDrain} channel, which is a channel with
two source ends through which it accepts data simultaneously and loses them
subsequently; (iv) the {\AsyncDrain} channel, which accepts data items
through only one of its two source channel ends at a time
and loses them; and (v)~the {\FIFO} channel, which is an asynchronous
channel with a buffer of capacity one. Additionally, there are
channels for data manipulation. For instance, the {\Filter} channel
always accepts a data item at its source end and synchronously passes
or loses it depending on whether or not the data item matches a
certain predefined pattern or data constraint. Finally, the
{\Transform} channel applies a user-defined function to the data item
received at its source end and synchronously yields the result at its
sink end.

Channels can be joined together using nodes. A node can be a \emph{source}, a
\emph{sink} or a \emph{mixed} node, depending on whether all of its coinciding
channel ends are source ends, sink ends or a combination of
both. Source and sink nodes together form the boundary nodes of a
connector, allowing interaction with its environment. Source nodes act
as synchronous replicators, and sink nodes as non-deterministic
mergers. A mixed node combines these two behaviors by atomically
consuming a data item from one of its sink ends at the time and
replicating it to all of its source ends.

\subsection{Automata-based Semantics for Reo}

The most basic model expressing formally the semantics of Reo is
constraint automata~\cite{BSA+06}. Transitions in a constraint
automaton are labeled with sets of ports that fire synchronously, as
well as with data constraints on these ports. The constraint
automata-based semantics for Reo is compositional, meaning that the
behavior of a complex Reo circuit can be obtained from the semantics
of its constituent parts using the product operator. Furthermore, the
hiding operator can be used to abstract from unnecessary details such
as dataflow on the internal ports of a connector.

\begin{definition}[Constraint automaton (CA)]
  \label{def:ca}
  A constraint automaton $\A=(S,\N,\rightarrow,s_0)$ consists of a set
  of states~$S$, a set of port names~$\N$, a transition relation
  $\mathord{\rightarrow} \subseteq S \times 2^{\N} \times DC \times
  S$, where $DC$ is the set of data constraints over a finite data
  domain $\mathit{Data}$, and an initial state $s_0 \in S$.
\end{definition}

\begin{figure}[t]
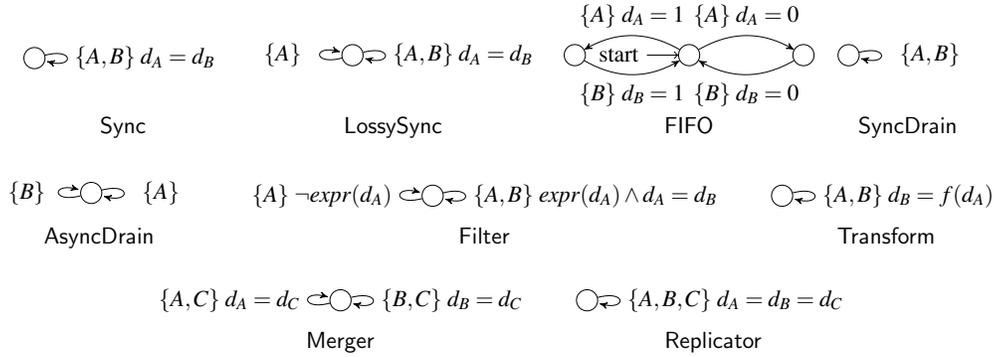

 \centering
 \scalebox{0.8}{
 \begin{tabular}{c}
     \begin{tabular}{c@{\quad}c@{\quad}c@{\quad}c}
         \tikz{
           \node[state] (q) {};
           \path[transition] (q) edge [loop right] node {
             \U{A,B}{d_A=d_B} } (q); } &
         \tikz{
           \node[state] (q) {};
           \path[transition] (q) edge [loop right] node { \U{A,B}{d_A=d_B} } (q)
           (q) edge [loop left] node { \TT{A}{} } (q); } &
         \raisebox{-6mm}{\tikz{
            \node[state,initial] (q) {};
            \node[state,right of=q] (p0) {};
            \node[state,left of=q] (p1) {};
            \path[transition] (q)  edge [bend left]  node[above] {
              \U{A}{d_A=0} } (p0)
            (p0) edge [bend left]  node[below] { \U{B}{d_B=0} } (q)
            (q)  edge [bend right] node[above] { \U{A}{d_A=1} } (p1)
            (p1) edge [bend right] node[below] { \U{B}{d_B=1} } (q); }} &
        \tikz{
           \node[state] (q) {};
           \path[transition] (q) edge [loop right] node { \T{A,B}{} } (q); } \\
        \Sync & \LossySync & \FIFO & \SyncDrain
      \end{tabular}
  \\
  \\
      \begin{tabular}{c@{\qquad}c@{\qquad}c}
        \tikz{
           \node[state] (q) {};
           \path[transition] (q) edge [loop right] node { \T{A}{} } (q)
           (q) edge [loop left] node { \U{B}{} } (q); } &
         \tikz{
        	\node[state] (q) {};
          	\path[transition] (q) edge [loop right] node {
              \U{A,B}{\mathit{expr}(d_A) \wedge d_A=d_B} }
          	(q) (q) edge [loop left] node { \U{A}{\neg \mathit{expr}(d_A)}
            } (q); } &
          \tikz{
            \node[state] (q) {};
            \path[transition] (q) edge [loop right] node {
              \U{A,B}{d_B=f(d_A)} } (q); }
        \\
        \AsyncDrain & \Filter & \Transform
      \end{tabular}
  \\
  \\
      \begin{tabular}{c@{\qquad}c}
          \tikz{
        	\node[state] (q) {};
          	\path[transition] (q) edge [loop left] node { \U{A,C}{d_A=d_C} } (q)
          	(q) edge [loop right] node { \U{B,C}{d_B=d_C} } (q); } &
          \tikz{
        	\node[state] (q) {};
          	\path[transition] (q) edge [loop right] node {
              \U{A,B,C}{d_A=d_B=d_C} } (q); }
        \\
        \channel{Merger} & \channel{Replicator}
      \end{tabular}
  \end{tabular}
  }
\caption{Constraint automata for basic Reo channels and nodes}
\label{fig:basicCA}
\end{figure}

\noindent
We write $\transition{q}{N,g}{p}$ instead of $(q, N, g, p) \in
\mathord{\rightarrow}$. Figure~\ref{fig:basicCA} shows the constraint
automata for the basic Reo channels. Note that we use the
set $\mathit{Data} = \{0, 1\}$ as data domain for the {\FIFO} channel. The behavior of any Reo circuit
composed from these channels can be obtained by computing the product
of the corresponding automata.

Constraint automata in their basic form do not express all the information about Reo node
communication and fail to represent the behavior of e.g. context-dependent channels.
An elemental example of such channels is a {\LossySync} channel that loses a data item only if the
environment or subsequent channels are not ready to consume it, i.e., it needs the information about the states of other channels or services to decide locally what to do with its data input. To address this problem, several other semantic models for Reo were introduced.

In intentional automata~\cite{Cos10} we distinguish two sets of ports
in their transition labels, a request set and a firing set. The
request set models the context, i.e., the readiness of the channel
ports to accept/dispense data, while the firing set models the actual
flow of data through the circuit ports. Accounting for the requests
that have arrived but have not been fired yet introduces additional
states in the model. Due to this fact, intentional automata rapidly
become large and difficult to manipulate.

\begin{figure}
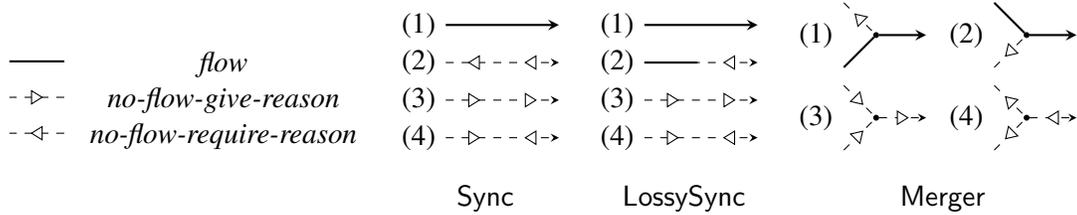

  \centering
 \begin{tabular}{c@{\hspace{0.8mm}}c}
    \tikz{\node(A){}; \node[left of=A,node distance=10mm](B){}; \draw [flow] (A) -- (B);} &
    \textit{flow} \\
    \tikz{\node(A){}; \node[left of=A,node distance=10mm](B){}; \draw [noflowRight] (A) -- (B);} &
    \textit{no-flow-give-reason} \\
    \tikz{\node(A){}; \node[left of=A,node distance=10mm](B){}; \draw [noflowLeft] (A) -- (B);} &
    \textit{no-flow-require-reason} 
 \end{tabular}
 \begin{tabular}{@{\hspace{1.6mm}}c@{\hspace{01.6mm}} c@{\hspace{1.6mm}} c}
	\tikz{
	\node (A1) {(1)};
	\node (B1) [right of=A1,node distance=20mm]{};
	\node (A2) [below of=A1,node distance=5mm] {(2)};
	\node (B2) [right of=A2,node distance=20mm]{};
	\node (A3) [below of=A2,node distance=5mm] {(3)};
	\node (B3) [right of=A3,node distance=20mm] {};
	\node (A4) [below of=A3,node distance=5mm] {(4)};
	\node (B4) [right of=A4,node distance=20mm] {};
	\node [below of=A4,node distance=3mm] {};
	\draw[sync,flow] (A1) to (B1);
	\draw[sync,noflowRightRight] (A2) to (B2);
	\draw[sync,noflowLeftLeft] (A3) to (B3);
	\draw[sync,noflowLeftRight] (A4) to (B4);
	}
	&
	\tikz{
	\node (A1) {(1)};
	\node (B1) [right of=A1,node distance=20mm]{};
	\node (A2) [below of=A1,node distance=5mm] {(2)};
	\node (C2) [right of=A2,node distance=12mm] {};
	\node (B2) [right of=A2,node distance=20mm]{};
	\node (A3) [below of=A2,node distance=5mm] {(3)};
	\node (B3) [right of=A3,node distance=20mm] {};
	\node (A4) [below of=A3,node distance=5mm] {(4)};
	\node (B4) [right of=A4,node distance=20mm] {};
	\node [below of=A4,node distance=3mm] {};
	\draw[sync,flow] (A1) to (B1);
	\draw[sync,noflow,postaction=decorate,
	      decoration={markings,mark=at position 0.75 with \noflowRight}] (A2) to (B2);
	\draw[sync,noflowLeftLeft] (A3) to (B3); \draw[sync,noflowLeftRight] (A4) to (B4);
	\draw[flow] (A2) to (C2);
	}
	&
	\tikz{
	\node (D1) {};
	\node[point] (C1) [left of=D1,node distance=8mm]{};
	\node (A1) [below left of=C1,node distance=8mm]{};
	\node (B1) [above left of=C1,node distance=8mm]{};
	\node[left of=C1,node distance=8mm]{(1)};
	\draw[sync,flow] (C1) to (D1);
	\draw[flow] (A1) to (C1);
	\draw[flow,noflowRight] (B1) to (C1);
	\node[right of=D1,node distance=20mm] (D2) {};
	\node[point] (C2) [left of=D2,node distance=8mm]{};
	\node (A2) [below left of=C2,node distance=8mm]{};
	\node (B2) [above left of=C2,node distance=8mm]{};
	\node[left of=C2,node distance=8mm]{(2)};
	\draw[sync,flow] (C2) to (D2);
	\draw[flow,noflowRight] (A2) to (C2);
	\draw[flow] (B2) to (C2);
	\node[below of=D1,node distance=11mm] (D3) {};
	\node[point] (C3) [left of=D3,node distance=8mm]{};
	\node (A3) [below left of=C3,node distance=8mm]{};
	\node (B3) [above left of=C3,node distance=8mm]{};
	\node[left of=C3,node distance=8mm]{(3)};
	\draw[sync,noflowLeft] (C3) to (D3);
	\draw[noflowLeft] (A3) to (C3);
	\draw[noflowLeft] (B3) to (C3);
	\node[below of=D2,node distance=11mm] (D4) {};
	\node[point] (C4) [left of=D4,node distance=8mm]{};
	\node (A4) [below left of=C4,node distance=8mm]{};
	\node (B4) [above left of=C4,node distance=8mm]{};
	\node[left of=C4,node distance=8mm]{(4)};
	\draw[sync,noflowRight] (C4) to (D4);
	\draw[noflowRight] (A4) to (C4);
	\draw[noflowRight] (B4) to (C4);
	}\\
    \Sync & \LossySync & \textsf{Merger} \\
  \end{tabular}
\caption{Examples of coloring semantics for Reo channels and nodes}
\label{fig:colorings}
\end{figure}

Connector coloring~\cite{CCA07} describes the behavior of Reo in a
compositional fashion by coloring the parts of the circuit using different colors that match on connected ports.
The basic idea in this model is to associate \emph{flow} and \emph{no-flow} colors to
channel ends. When three colors are used, the model captures context-dependent
behavior by propagating negative information about the exclusion of
dataflow through the connector. This model is used currently as a
theoretical basis for Reo circuit animation and simulation tools. Figure~\ref{fig:colorings} shows examples of coloring semantics for basic Reo channels and connectors.

In action constraint automata~(ACA)~\cite{KCA10}, we distinguish several kinds of actions triggered on channel ports to
signal the state changes of the channel. Formally, ACA can be defined as follows:
\begin{definition}[Action constraint automaton (ACA)]
  \label{def:dca}
  An action constraint automaton $\A=(S,\N,$ ${\rightarrow},s_0)$
  consists of a set of states~$S$, a set of action names~$\N$ derived
  from a set of port names~$\M$ and a set of admissible action types
  $\calT$, a transition relation $\mathord{\rightarrow} \subseteq S
  \times 2^{\N} \times \mathit{DC} \times S$, where $\mathit{DC}$ is
  the set of data constraints over a finite data domain
  $\mathit{Data}$, and an initial state $s_0 \in S$.
\end{definition}

\noindent
We introduce an injective function $\mathit{act} : \M \times \calT
\rightarrow {\N} $ to define action names for each pair of a port name
and an action type observed on the port.  For example, the function
$\mathit{act}(m, \alpha) = \alpha \cdot m$, for $ m \in \M, \alpha \in
\calT$, where `$\cdot$' is a standard concatenation operator, can be
used to obtain a set of unique action names given sets of distinctive
Reo port names and types of observable actions. This model can be used, e.g., to represent a sequential data flow within a synchronous region and account for time delays in synchronous channels by distinguishing port blocking and unblocking events as well as the start and the end of data transfer through a port.
Coloring semantics can also be represented in a form of ACA using three actions to convey the possibility of data \emph{flow} as well as \emph{requiring} and \emph{giving reasons} for \emph{no-flow}.

\subsection{Process algebra-based Semantics for Reo}
\label{sect:mcrl2}

In our recent work~\cite{KKV11}, we represented the aforementioned semantic models for Reo using the process algebra~{\mCRL}.
This allowed us to apply a set of verification tools developed for this specification language to analyze Reo connectors.

The basic notion in {\mCRL} is the action. Actions represent atomic
events and can be parameterized with data. Actions in {\mCRL} can be
synchronized. In this case, we speak of multiactions which are
constructed from other actions or multiactions using the so-called
synchronization operator~$|$, such as the multiaction
$a|b|c$ of simultaneously performing the actions $a$, $b$
and~$c$. The synchronization operator is commutative, i.e., multiactions $a|b$ and $b|a$ are equivalent. The special action~$\tau$ (tau) is used to refer to an internal, unobservable action. Processes are defined by process expressions, which are
compositions of actions and
multiactions using a number of operators. Among the basic operators are the following:
  (i) \emph{deadlock} or \emph{inaction} $\delta$, which does not
    display any behavior;
  (ii) \emph{alternative composition}, written as $p \mkern2mu +
    \mkern2mu q$, which represents a non-deterministic choice between
    the processes $p$ and~$q$;
  (iii) \emph{sequential composition}, written $p \cdot q$, which
    means that $q$ is executed after $p$, assuming that $p$
    terminates;
  (iv) the \emph{conditional operator} or the \emph{if-then-else}
    construct, written as $c \rightarrow p \diamond q$, where $c$ is a
    data expression that evaluates to true or false;
  (v) \emph{summation} $\Sigma_{d:D} \:\, p$ where $p$ is a process
    expression in which the data variable~$d$ may occur, used to
    quantify over a data domain~$D$;
  (vi) \emph{parallel composition} or \emph{merge} $p \mergeProc q$,
    which interleaves and synchronizes the multiactions of~$p$ with
    those of~$q$, where synchronization is governed by a
    communication function (see below);
  (vii) \emph{allow} $\nabla_V(p)$, where only actions in~$p$ from the
  set~$V$ are allowed to occur;
  (viii) the \emph{encapsulation} $\partial_H (p)$, where $H$ is a set
    of action names that are not allowed to occur;
  (ix) the \emph{renaming operator} $\rho_R(p)$, where $R$ is a set
    of renamings of the form $a \mathbin{\rightarrow} b$, meaning that
    every occurrence of the action~$a$ in~$p$ is replaced by the
    action~$b$;
  (x) the \emph{communication operator} $\Gamma_C(p)$, where $C$ is
    a set of communications of the form $a_0 |...|a_n
    \mathbin{\mapsto} c$, which means that every group of actions $a_0
    |...|a_n$ within a multiaction is replaced by the action~$c$;
  (xi) \emph{hiding } $\tau_I(p)$, which renames all actions in~$I$
  of~$p$ into $\tau$.
It is possible to define recursive processes in {\mCRL}. However, allow, encapsulation, hiding and communication operators can not be used within recursive processes. Structured operational semantics for the aforementioned {\mCRL} operators can be found in~\cite{Gro07:dagstuhl}.

The {\mCRL} language provides a number of built-in datatypes (e.g., boolean, natural, integer) with a set of usual arithmetic operations.  Moreover, an arbitrary structured type in {\mCRL}
can be declared by a construct of the form
\begin{displaymath}
  \begin{array}{@{}l}
     \textbf{sort} \, S  = \textbf{struct} \ c_1( \mkern2mu
     p_1^1 {:} \mkern1mu S_1^1,
     \mkern4mu \ldots , \mkern2mu
     p_1^{\mkern1mu k1} {:} \mkern1mu S_1^{\mkern1mu k1} \mkern2mu )
     \mkern1mu ? \mkern1mu r_1
     \mkern4mu \mid \mkern4mu \ldots \mkern4mu \mid \mkern4mu
     c_n( \mkern1mu
     p_n^1 {:} \mkern1mu S_n^1,
     \mkern4mu \ldots , \mkern2mu
     p_n^{\mkern1mu kn} {:} \mkern1mu S_n^{\mkern1mu kn} \mkern2mu )
     \mkern1mu ? \mkern1mu r_n \, ;  \\
  \end{array}
\end{displaymath}
This construct defines the type $S$ together with constructors $c_i \colon S_i^1
\times \ldots \times S_i^{\mkern1mu ki} \to S$, projections
$p_i^{\mkern1mu j} \colon S \to S_i^{\mkern1mu j}$, and type
recognition functions $r_i \colon S \to \Bool$.

The {\mCRL} toolset allows users to verify software models specified in the {\mCRL} language. It includes a tool for
converting {\mCRL} specifications into linear form (a compact symbolic representation of the corresponding LTS), a tool for generating explicit LTSs from linear process specifications (LPS), tools for optimizing and visualizing these LTSs, and many other useful facilities. A detailed overview of the available software can be found at the {\mCRL} web site\footnote{\url{http://www.mcrl2.org/}}.

\begin{table}[t]
\centering
\caption{{\normalfont\mCRL} encoding for channels and nodes: CA semantics}
\begin{tabular}{|@{\quad}r@{$\;=\;$}l @{\quad}|}
  \hline
  $\Sync$\rule{0pt}{10pt}
    & $\Sigma_{d{:}\Data} \,\, A(d)|B(d) \cdot \Sync$ \\
  $\LossySync$  & $\Sigma_{d{:}\Data} \,\, (A(d)|B(d) + A(d)) \cdot \LossySync$ \\
  $\SyncDrain$  & $\Sigma_{d_1,d_2: \Data} \,\, A(d_1)|B(d_2) \cdot \SyncDrain$ \\
  $\AsyncDrain$ & $\Sigma_{d{:} Data} \,\, (A(d) + B(d)) \cdot \AsyncDrain$ \\
  $\FIFO(f:\DataFIFO)$ & $\Sigma_{d:Data}$ \\
        \multicolumn{2}{|r|}{
  		$\qquad\qquad\qquad\qquad
  		(\mathit{isEmpty}(f) \to A(d) \cdot \FIFO(\mathit{full}(d)) \diamond \,
  		B(e(f)) \cdot \FIFO(\mathit{empty})) \quad $}\\
  $\Filter$ & $\Sigma_{d{:}Data} \,\, (\mathit{expr}(d) \to A(d)|B(d)
  \diamond A(d)) \cdot \Filter$ \\
  $\Transform$ & $\Sigma_{d{:} Data} \,\,  A(d)|B(f(d) ) \cdot  \Transform$ \\
  \hline \hline
  $\channel{Merger}$ & $\Sigma_{d{:}Data} \,\, (A(d)|C(d) + B(d)|C(d))
  \cdot \channel{Merger}$ \\
  $\channel{Replicator}$ & $\Sigma_{d{:}Data} \,\, A(d)|B(d)|C(d)
  \cdot \channel{Replicator}$\\
  $\channel{Router}$ & $\Sigma_{d:Data} \,\, (A(d)|B(d) + A(d)|C(d))
   \cdot \channel{Router}$\\
 \hline
\end{tabular}
\label{tab:encodings}
\end{table}

The presence of multiactions in {\mCRL} makes it possible to compositionally map Reo to process specifications and compose a connector by synchronizing actions on joint ports. Thus, {\mCRL} models for Reo circuits are generated in the following way: observable events,
i.e., data flow on the channel ends, are represented as atomic
actions, while data items observed at these ports are modeled as
parameters of these actions. Analogously, we introduce a process for
every node and actions for all channel ends meeting at the node. The
encodings for the basic Reo channels and nodes are listed in
Table~\ref{tab:encodings}. Given process definitions for all channels and nodes, a composite process
that models a complete Reo connector is built by forming a parallel
composition of these processes and synchronizing actions for
coinciding node/channel ends. Node/channel end synchronization is
enforced using the {\mCRL} operators communication and encapsulation.
For example, an {\mCRL} process for the replicator circuit
in Figure~\ref{fig:basicChannels} can be formed from three synchronous
channels
\begin{displaymath}
    \Sync1 = A|X_1 \mydot \Sync1 \, , \quad
    \Sync2 = Y_1|B \mydot \Sync2 \, , \quad
    \Sync3 = Z_1|C \mydot \Sync3
\end{displaymath}
and a replicator node
\begin{displaymath}
  \ReplicatorNode = X_2|Y_2|Z_2 \mydot \ReplicatorNode
\end{displaymath}
applying the communication and blocking operators to their parallel
composition:
\begin{displaymath}
  \begin{array}{@{}l}
    \channel{ReplicatorCircuit} =
    \partial_{\{X_1,Y_1,Z_1,X_2,Y_2,Z_2\}}\bigl (
    \Gamma_{\{X_1|X_2\rightarrow \tau, Y_1|Y_2 \rightarrow \tau, Z_1|Z_2
      \rightarrow \tau \}}(
    \\\qquad \qquad \qquad
    \Sync1 \parallel \Sync2 \parallel \Sync3 \parallel
    \ReplicatorNode
    \, ) \, \bigr );
\end{array}
\end{displaymath}
Here we assume that the sink end~$X_1$ of the channel $\Sync1$ is
connected to the source end~$X_2$ of the node $\ReplicatorNode$, while
sink ends $Y_2$ and~$Z_2$ of the node are connected to source ends $Y_1$
and~$Z_1$ of channels $\Sync2$ and $\Sync3$. Optionally, the {\mCRL} hiding operator can be employed for
abstracting the flow in selected nodes. For simplicity, we omitted the encoding of data parameters in this example.

For the treatment of data we assume, in the context of a given
connector, a global datatype given as the custom sort {\Data}. Given such a datatype, we can use the {\mCRL} summation
operator to define data dependencies imposed by channels. For the
{\FIFO} channel we additionally define the datatype
\begin{displaymath}
  \begin{array}{@{}l}
    \textbf{sort} \,\, \DataFIFO = \struct \;
    \mathit{empty}?\mathit{isEmpty} \mid \mathit{full}(e{:}\Data)?
    \mathit{isFull}
  \end{array}
\end{displaymath}
which allows us to specify whether the buffer of the {\FIFO} channel
is empty or full, and if it is full, what value is stored in it.
Additionally, we introduce a special kind of node,
{\JoinNode}, which synchronizes all ends of incoming channels, forms a tuple of data
items received and replicates it to the source ends of all outgoing
channels. More details on data handling in Reo and {\mCRL} can be found in~\cite{KKV11}.

Table~\ref{tab:encodings-aca} shows the {\mCRL} encodings for the basic Reo channels and nodes according to the ACA model with four actions: \emph{block} and \emph{unblock} actions are used to establish port communication within a single transaction and release channel ports involved in such a transaction, respectively. The \emph{start} and \emph{finish} actions are used to represent the start and the end of dataflow through a blocked channel port. In our encoding, we use prefix letters $b$, $u$, $s$ and $f$ in front of Reo port names to denote block, unblock, start and finish actions observed on these ports. Since data support in the new translation is analogous to the case of the CA-based translation, we omit its discussion here and for simplicity show only the data-agnostic mapping. As in the CA approach, we construct nodes compositionally. Given process definitions for all channels and nodes, a composite process that models the complete Reo connector is built by forming a parallel composition of these processes and synchronizing communicating actions for the coinciding node/channel ends.

\begin{table}[t]
\centering
\caption{{\normalfont\mCRL} encoding for channels and nodes: ACA semantics}
\label{tab:encodings-aca}
   \begin{tabular}{|r@{$\;=\;$}l|}
      \hline
      $\Sync$\rule{0pt}{10pt}
        & $bA|bB \cdot sA|sB \cdot fA|fB \cdot uA|uB \cdot \Sync$ \\
      $\LossySync$ & $(bA|bB \cdot sA|sB \cdot fA|fB \cdot uA|uB + bA \cdot sA \cdot fA \cdot uA) \cdot \LossySync$ \\
      $\SyncDrain$ & $bA|bB \cdot ( $\\
      \multicolumn{2}{|l|}{\qquad \qquad \qquad $sA \cdot (sB \cdot (fA \cdot fB + fB \cdot fA + fA|fB) + fA \cdot sB \cdot fB + sB|fA \cdot fB)+$}\\
      \multicolumn{2}{|l|}{\qquad \qquad \qquad $sB \cdot (sA \cdot (fA \cdot fB + fB \cdot fA + fA|fB) + fB \cdot sA \cdot fA + sA|fB \cdot fA) + $}\\
      \multicolumn{2}{|l|}{\qquad \qquad \qquad $sA|sB \cdot (fA \cdot fB + fB \cdot fA + fA|fB)) \cdot uA|uB \cdot \SyncDrain$ }\\
      $\AsyncDrain$ & $(bA \cdot sA \cdot fA \cdot uA + bB \cdot sB \cdot fB \cdot uB) \cdot \AsyncDrain$ \\
      $\FIFO(f:\DataFIFO)$ & $\mathit{isEmpty}(f) \to bA \cdot sA \cdot fA \cdot uA \cdot \FIFO(\mathit{full}) \, \diamond $ \\
      \multicolumn{1}{|l}{} & \qquad \qquad \qquad $bB \cdot sB \cdot fB \cdot uB \cdot \FIFO(\mathit{empty})$\\
      \hline \hline
      $\channel{Merger}$ & $(bA|bC \cdot sA|sC|fA|fC.uA|uC + $\\
      \multicolumn{1}{|l}{} & \qquad $bB|bC \cdot sB|sC|fB|fC \cdot uB|uC) \cdot \channel{Merger}$ \\
      $\channel{Replicator}$ & $bA|bB|bC \cdot sA|sB|sC \cdot fA|fB|fC \cdot uA|uB|uC \cdot \channel{Replicator}$\\
      \hline
   \end{tabular}
\end{table}

To incorporate the colorings in our encoding in {\mCRL}, we represent colors as data
parameters of actions~\cite{KKV11}. However, since the summation over a finite domain in {\mCRL} is just an alternative choice of the same action with various parameters,
we can represent every parameterized action as an alternative choice of several non-parameterized actions. This allows us to represent coloring semantics as shown in Table~\ref{tab:encodings-colouring}.
For every port $X,$ we consider three actions, $wX, \, rX$ and $gX$ which are abbreviations for actions \emph{flow}, \emph{no-flow-require-reason}, and \emph{no-flow-give-reason} observations on channel ports. The advantage of this approach over the use of parameterized actions is the possibility to hide no-flow labels.

\begin{table}[t]
\centering
\caption{{\normalfont\mCRL} encoding for channels and nodes: coloring semantics}
\label{tab:encodings-colouring}
   \begin{tabular}{|r@{$\;=\;$}l|}
      \hline
      $\Sync$ & $(wA | wB \choice rA | gB \choice gA | rB \choice gA | gB) \seq \Sync$ \\
      $\LossySync$ & $(wA | wB \choice wA | gB \choice gA | rB \choice gA | gB) \seq \LossySync$ \\
      $\SyncDrain$ & $(wA | wB \choice rA | gB \choice gA | rB \choice gA | gB) \seq \SyncDrain$ \\
      $\AsyncDrain$ & $(wA | gB \choice gA | wB \choice rA | wB \choice rB | wA  \choice gA | gB ) \cdot \AsyncDrain$ \\
      $\FIFO(f:\DataFIFO)$ & $\mathit{isEmpty}(f) \to ((wA | rB + wA | gB) \seq \FIFO(\mathit{full}) \choice  $\\
      \multicolumn{1}{|l}{} & \qquad \qquad $(gA|rB + gA|gB) \seq \FIFO(\mathit{empty})) \, \diamond \,$ \\
      \multicolumn{1}{|l}{} & \qquad \qquad $((rA | wB + gA | wB) \seq \FIFO(\mathit{empty}) \choice $\\
      \multicolumn{1}{|l}{} & \qquad \qquad $(rA | gB + gA | gB) \seq \FIFO(\mathit{full})) $\\
      \hline \hline
      $\channel{Merger}$ & $wA | gB | wC \choice gA | wB | wC \choice rA | rB | gC \choice gA | gB | rC) \seq \channel{Merger} $ \\

      $\channel{Replicator}$ & $(wA | wB | wC + rA | rB | gC   + rA | gB | rC  + gA  | gB | gC) \seq \channel{Replicator}$\\
      \hline
   \end{tabular}
\end{table}

Thus, the process algebra {\mCRL} provides a common ground for expressing most important semantic models for Reo preserving their compositionality.

\section{Input-output Conformance Testing}
\label{sect:testingTheory}

In this section, we briefly introduce a model-based test generation theory for testing input-output conformance (\emph{ioco}) of an implementation and a given specification~\cite{Tre08}. Transition labels in (action) constraint automata represent sets of simultaneously observable actions on Reo ports with enabling guards while in the original definitions on LTS each transition refers to a single observable action. As follows from our mapping of constraint-automata-based semantics of Reo to LTS, each set of transition labels $\{A,B,C\}$ in a CA corresponds to a transition with a unique action label $A|B|C$ in the corresponding LTS, which further can be renamed to an action $ABC$. Assuming that the semantics of Reo is given in a form of such LTS, we can apply the \emph{ioco} testing theory to test Reo. In the following we redefine all necessary concepts of the \emph{ioco} testing theory using CA, the original definitions on LTS can be found in~\cite{Tre08}.

Let $L^*$ be the set of all finite sequences over a set $L$ and $\epsilon$ denote the empty sequence. Given finite sequences $\sigma_1$ and $\sigma_2$, we denote their concatenation $\sigma_1 \seq \sigma_2$. If for some automaton there exists a trace $\longtransition{q}{N_1 \seq \tau \seq \tau \seq N_2 \seq \tau \seq N_3 \seq \tau}{p},$ where $ N_1,N_2,N_3 \in L$ are sets of actions representing constraint automata labels and $\tau$ is a special action that refers to any set of unobservable constraint automata ports, we write $\trace{p}{N_1 \seq N_2 \seq N_3}{q}$ for the $\tau-$abstracted sequence of observable actions and say that $p$ is able to perform the trace $N_1 \cdot N_2 \cdot N_3 \in L^*.$ As we demonstrated in~\cite{KKV10:fmco}, every state $s$ of a CA can be identified with a behaviorally equivalent {\mCRL} process $p$. We exploit this correspondence in the rest of the paper and do not distinguish between CA states and processes associated with these states. The following definitions are needed to formally define the \emph{ioco} testing relation for a given specification and a system implementation.

\begin{definition}
  Let $p$ be a process associated with the initial state $s_0$ of a constraint automaton $\A=(S,\N,\rightarrow,s_0)$ and $\sigma \in L^*$ where  $L = 2^{\N} \times DC$ is a set of the constraint automaton labels.
 \begin{enumerate}
   \item $init(p) = \{\rho \in L \cup \tau \, | \, \transition{p}{\rho}{} \}.$
   \item $traces(p) = \{ \sigma \in L^* \, | \, \trace{p}{\sigma}{} \}$
   \item $p \, \mbox{\bf after} \, \sigma = \{ p'\, | \, \trace{p}{\sigma}{p'} \} $
   \item $P \, \mbox{\bf after} \, \sigma = \bigcup {p \, \mbox{\bf after} \, \sigma \, | \, p \in P},$ where $P \subseteq S$ is a set of states.
   \item $P \, \mbox{\bf refuses} \, A = \exists p \in P, \, \forall \rho \in A \cup \tau \, : \, \notransition{p}{\rho}{},$ where $P\subseteq S$ and $A \subseteq L$.
   \item $der(p) = \{p' \, | \, \exists \sigma \in L^* \, : \, \trace{p}{\sigma}{p'} \}$
   \item $p$ has \emph{finite behavior} if there is a natural number $n$ such that all traces in $traces(p)$ have length smaller than $n$.
   \item $p$ is a \emph{finite state} if the number of reachable states $der(p)$ is finite.
   \item $p$ is \emph{deterministic} if, for all $\sigma \in L^*,\, p \, \mbox{\bf after} \, \sigma $ has at most one element.
   \item $p$ is \emph{image finite} if, for all $\sigma \in L^*,\, p \, \mbox{\bf after} \, \sigma $is finite.
   \item $p$ is \emph{strongly convergent} if there is no state of $p$ that can perform an infinite sequence of internal transitions.
   \item $\mathcal{CA}(L)$ is the class of image finite and strongly convergent constraint automata with labels in $L$.
 \end{enumerate}
\end{definition}

\begin{definition}[Constraint automaton with Inputs and Outputs]
  \label{def:ca}
  A \emph{constraint automaton with inputs and outputs} is a constraint automaton $\A=(S,\N,\rightarrow,s_0) \in \mathcal{CA}(L_I \cup L_U)$, where
  $L_I$ and $L_U,$ $L_I \cap L_U = \emptyset$ are countable sets of disjoint input and output labels.
\end{definition}

LTS with inputs and outputs are used as formal specifications for \emph{ioco} testing theory. Being a variant of LTS, constraint automata with inputs and outputs are used in our work to represent system-under-test specifications. This does not mean that these specifications have to be written explicitly in a form of automata: it suffices that a specification language, e.g., Reo, had semantics expressed in the form of constraint automata with inputs and outputs.

\begin{definition}[Input-Output Constraint Automaton]
  \label{def:ioca}
  An \emph{input/output constraint automaton} (IOCA) is a constraint automaton with inputs and outputs $\A=(S,\N,\rightarrow,s_0)$ where all inputs are enabled in any reachable state, i.e., $\forall s \in der (s_0), \,  \forall N \subseteq L_I: \trace{s}{N}{}$.
\end{definition}
Let $\mathcal{CA}(L_I, L_U)$ denote the class of all constraint automata with inputs in $L_I$ and outputs in $L_U$.
The class of input-output constraint automata with inputs in $L_I$ and outputs in $L_U$ is denoted by $\mathcal{IOCA}(L_I, L_U) \subseteq \mathcal{CA}(L_I, L_U).$
A constraint automaton with inputs and outputs can be converted to an input-output constraint automaton by adding a self-loop transition with labels from $L_I$ to every reachable state. This operation is called \emph{angelic completion}~\cite{Tre08}. Input-output constraint automata are used to model systems in which inputs are initiated by the environment and never refused by the system and outputs are initiated by the system and never refused by the environment. The input enabledness of system implementations is required in \emph{ioco} testing theory to define the relation between the inputs generated by the tester and the observable outputs.

Since input-output constraint automata are just a particular type of constraint automata, all definitions for the latter apply, including the definitions of product and hiding operations. A state $q$ of a process $p$ without output actions, i.e., $\forall \rho \in L_U \, | \, \notransition{q}{\rho}{},$  is called \emph{suspended} or \emph{quiescent} and is denoted $\delta(q)$. The external observer of a system in a quiescent state does not see any outputs. Such a situation with no observations can be considered as a special action, denoted as $\delta$. In our test cases, we allow system transitions $\transition{p}{\delta}{}$ meaning that $p$ cannot perform any output actions. It is also possible to extend traces with $\delta$, e.g., $\trace{p}{N_1 \cdot \delta \cdot N_2 \cdot N_3}{}$, where $N_1,N_2 \in L_I, \, N_3 \in L_U,$ expresses the fact that after the input $N_1$ was observed, the system remained quiescent, while after the input $N_2$, the system produced output $N_3.$ The \emph{quiescent traces} of $p$ are those that may lead to quiescent states, i.e., $$Qtraces(p) = \{\sigma \in L^* | \exists p' \in (p \, \mbox{\bf after} \, \sigma) : \, \delta(p')\}.$$
Traces that may contain the quiescence action are called \emph{suspension traces}. More formally, the suspension traces are
$$Straces(p) = \{\sigma \in L^*_{\delta} \, | \, \transition{p_{\delta}}{\sigma}{}\},$$ where $L_{\delta} = L \cup \delta$ and $p_{\delta}$ is a process defined by a constraint automaton $\A=(S,\N,\rightarrow,s_0)$ with inputs $L_I$, outputs $L_U \cup \delta$ and a transition relation $\rightarrow \cup \rightarrow_{\delta},$ such that $\rightarrow_{\delta} = \{\transition{s}{\delta}{s} \, | \, s \in S, \delta(s)\}.$

To test a system using the \emph{ioco} testing theory, we assume that a tester is an environment which is able to provide inputs and observe system outputs including quiescence. This environment must be able to accept any output produced by the system. Thus, the behavior of a tester can be modeled as IOCA with inputs and outputs exchanged.
The occurrence of a special symbol $\theta \notin L_I \cup L_U \cup \tau \cup \delta $ in tests indicates the detection of quiescence. Practically this means that the tester has to wait for a certain time-out to conclude that the system did not produce an output.
Since test case execution must always lead to a verdict, we include two special states reachable from any other state of a testing IOCA: {\bf fail}, {\bf pass} $\in S$.
Thus, a test case is defined as follows in \emph{ioco}:

\begin{definition}[Test case]
  \label{def:test-case}
   A \emph{test case} $t$ for an implementation with inputs in $L_I$ and outputs in $L_U$ is an IOCA $\A=(S,\N,\rightarrow,s_0) \in \mathcal{IOCA}(L_I, L_U \cup \theta)$ such that
   \begin{itemize}
     \item $t$ is finite and deterministic;
     \item $S$ contains two special states {\bf pass} and {\bf fail}, {\bf pass} $\neq$ {\bf fail};
     \item $t$ has no cycles except those in states {\bf pass} and {\bf fail};
     \item $\forall s \in S$ it holds that $init(s) = {a}\cup L_U \, | \, a \in L_I$ or $init(s) = L_U \cup \theta$.
   \end{itemize}
\end{definition}

The class of test cases for implementations with inputs in $L_I$ and outputs in $L_U$ is denoted $\mathcal{TTS}(L_U, L_I).$
A run of a test case $t \in \mathcal{TTS}(L_U, L_I)$ with an implementation under test $i \in \mathcal{IOCA}(L_I, L_U)$ corresponds to the parallel synchronization of behavior expressed by the tester and the system. However, the usual parallel synchronization needs to be extended to account for special labels $\delta$ and $\theta.$ Such an extension, denoted by $t \rceil | i$, is defined by the following inference rules:
\begin{displaymath}
 {\large
 \begin{array}{ccc}
 \frac{\transition{i}{\tau}{i'\,}}{\transition{\, t \rceil | i}{\tau}{t \rceil | i'}} \quad & 
 \frac{\transition{t}{a}{t'}, \quad \transition{i}{a}{i'}}{\transition{t \rceil | i}{a}{t \rceil | i'}} \quad &
 \frac{\transition{t}{\theta}{t'}, \quad \transition{i}{\delta}{}}{\transition{\, t \rceil | i}{\theta}{t \rceil | i'}}.  \\
\end{array} 
}
\end{displaymath}
Here $a \in L_I \cup L_U.$ The resulting system runs without deadlocks. This property follows immediately from the definition of test cases: since $\forall s \in S$ it holds that $init(s) = {a} \cup L_U \, | \, a \in L_I$ or $init(s) = L_U \cup \theta$, we can conclude that either an action $a$ can always be performed on the implementation or $i$ produces some output $x \in L_U \cup \theta.$

\begin{definition}[\emph{Ioco} relation]
   \label{def:ioco}
   Given a set of inputs $L_I$ and a set of outputs $L_U,$ the relation $\mbox{\bf ioco} \subseteq \mathcal{IOCA}(L_I, L_U) \times \mathcal{CA}(L_I, L_U)$ is defined as follows:
     $$i \, \mbox{\bf ioco} \, s = \forall \sigma \in Straces(s): \, out(i \, \mbox{\bf after} \, \sigma) \subseteq out (s \, \mbox{\bf after} \, \sigma )$$ where
     for any state $s$ of a CA $out(s) = \{x \in L_U \, | \, \transition{s}{x}{}\} \cup \delta \, | \, \delta(s)$ and for a set of states $S$ $out(S) = \cup \{out(s) \, | \, s \in S\}$
\end{definition}

For more details about \emph{ioco} testing theory, i.e., test generation algorithm and the analysis of its coverage, refer to~\cite{Tre08}. The extension of \emph{ioco} to test component-based systems is presented in~\cite{FGG07}. Aichernig and Weiglhofer propose a unification of \emph{ioco} relation by lifting the definition from LTS to reactive processes. In the rest of this paper, we discuss the application of the presented testing theory to detect errors in implementations of Reo coordination protocols. Given a Reo circuit specification, we use the \emph{ioco}-based test generation algorithm to produce sets of inputs and judge the correctness of the implementations by observing its outputs. Inputs in our approach essentially represent sets of boundary ports of the circuit ready to accept data items while outputs are actual observations of dataflow on these ports.

\section{Testing Channel-based Service Connectors}
\label{sect:testingReo}

\begin{figure}
 \centering
 \scalebox{0.9}{
    \begin{tabular}{cc}
    \subfigure[Specification]{
        \begin{tabular}{cc}
            \begin{tikzpicture}[>=stealth]
              \node[inner sep=0, label=left:$A$] at (0, 0)   (A)   {};
              \node[point]  at (0.5, 0)    (X1) {};
              \node[point]  at  (2,0.5)  (X2) {};
              \node[point]  at  (2,-0.5) (X3) {};
              \node[inner sep=0, label=right:$B$] at (2.5,0.5) (B) {};
              \node[inner sep=0, label=right:$C$] at (2.5,-0.5)(C) {};
              \draw[sync] (A) to (X1);
              \draw[fifo] (X1) to (X2);
              \draw[fifo] (X1) to (X3);
              \draw[sync] (X2) to (B);
         	  \draw[sync] (X3) to (C);
            \end{tikzpicture} &
            \raisebox{-3pt}{
            \begin{tikzpicture}[->,>=stealth]
                  \tikzstyle{every state}=[draw=black,text=black,minimum size=10pt,node distance=1.6cm]
                  \node[state] (s1) {};
                  \node[state] (s2) [right of=s1] {};
                  \node[state] (s3) [above of=s2] {};
                  \node[state] (s4) [above of=s1] {};
                  \path (s1) edge [bend right] node[fill=white] {$\{A\}$} (s3)
                  (s3) edge [bend right] node [fill=white] {$\{B, C\}$} (s1)
                  (s3) edge node [right] {$\{B\}$} (s2)
                  (s2) edge node [below] {$\{C\}$} (s1)
                  (s3) edge node [above] {$\{C\}$} (s4)
                  (s4) edge node [left] {$\{B\}$} (s1);
            \end{tikzpicture}}
             \\
            Reo model & Constraint automaton
        \end{tabular}
    }
     &
    \subfigure[Implementation]{
        \begin{tabular}{cc}
            \begin{tikzpicture}[>=stealth]
              \node[inner sep=0, label=left:$A$] at (0, 0)   (A)   {};
              \node[point]  at (0.5, 0)    (X1) {};
              \node[point]  at  (2,0.5)  (X2) {};
              \node[point]  at  (2,-0.5) (X3) {};
              \node[inner sep=0, label=right:$B$] at (2.5,0.5) (B) {};
              \node[inner sep=0, label=right:$C$] at (2.5,-0.5)(C) {};
              \draw[sync] (A) to (X1);
              \draw[fifo] (X1) to (X2);
              \draw[fifo] (X1) to (X3);
              \draw[syncdrain] (X2) to (X3);
              \draw[sync] (X2) to (B);
         	  \draw[sync] (X3) to (C);
            \end{tikzpicture}
            &
            \raisebox{3pt}{
            \begin{tikzpicture}[->,>=stealth]
                  \tikzstyle{every state}=[draw=black,text=black,minimum size=10pt,node distance=1.6cm]
                  \node[state] (s1) {};
                  \node[state] (s2) [right of=s1] {};
                  \path (s1) edge [bend left] node[above] {$\{A\}$} (s2)
                  (s2) edge [bend left] node [below] {$\{B, C\}$} (s1);
            \end{tikzpicture}}
            \\
            Reo model & Constraint automaton
        \end{tabular}
   }
   \end{tabular}
}
\caption{Specifications of a Reo connector and its wrong implementation (Example 2)}
\label{fig:example2}
\end{figure}
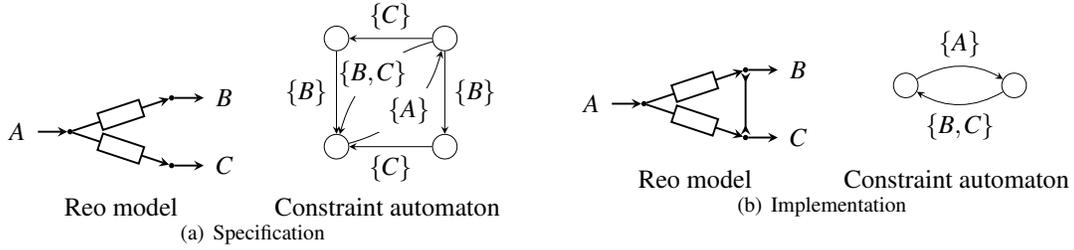

To enable testing of Reo connectors, we extend constraint automata with actions that represent input/output events.
Figure~\ref{fig:example1} shows a Reo connector specification and an erroneous implementation where {\Sync} and {\FIFO} channels are swapped. Figure~\ref{fig:example2} shows another sample specification and a wrong implementation where the {\SyncDrain} channel is erroneously added to the circuit. The goal of testing is to detect such errors automatically by providing inputs and observing outputs obtained from a wrong implementation which do not occur in the specification. Note that we use Reo to model both a specification and an erroneous implementation for illustration purposes only. In practice these errors may correspond to wrong implementation code such as e.g., wrong type of communication (synchronous vs. asynchronous) in the first example or wrongly enforced synchronization on two ports in the second example.

\begin{figure}
 \centering
 \scalebox{0.9}{
    \begin{tabular}{c@{\quad}c}
    \subfigure[Specification]{
        \begin{tabular}{c@{\quad}c}
            \begin{tikzpicture}[>=stealth]
                  \node[inner sep=0, label=left:$A$] at (0, 0)   (A)   {};
                  \node[point]  at (0.5, 0)    (X1) {};
                  \node[point]  at  (1.5,0.5)  (X2) {};
                  \node[point]  at  (1.5,-0.5) (X3) {};
                  \node[inner sep=0, label=right:$B$] at (2,0.5) (B) {};
                  \node[inner sep=0, label=right:$C$] at (2,-0.5)(C) {};
                  \draw[sync] (A) to (X1);
                  \draw[fifo] (X1) to (X2);
                  \draw[sync] (X1) to (X3);
                  \draw[sync] (X2) to (B);
             	  \draw[sync] (X3) to (C);
            \end{tikzpicture} &
            \raisebox{3pt}{
            \begin{tikzpicture}[->,>=stealth]
                  \tikzstyle{every state}=[draw=black,text=black,minimum size=10pt,node distance=1.6cm]
                  \node[state] (s1) {};
                  \node[state] (s2) [right of=s1] {};
                  \path (s1) edge [bend left] node[above] {$\{A, C\}$} (s2)
                  (s2) edge [bend left] node [below] {$\{B\}$} (s1);
            \end{tikzpicture}}
             \\
            Reo model & Constraint automaton
        \end{tabular}
    }
     &
    \subfigure[Implementation]{
        \begin{tabular}{c@{\quad}c}
            \begin{tikzpicture}[>=stealth]
                  \node[inner sep=0, label=left:$A$] at (0, 0)   (A)   {};
                  \node[point]  at (0.5, 0)    (X1) {};
                  \node[point]  at  (1.5,0.5)  (X2) {};
                  \node[point]  at  (1.5,-0.5) (X3) {};
                  \node[inner sep=0, label=right:$B$] at (2,0.5) (B) {};
                  \node[inner sep=0, label=right:$C$] at (2,-0.5)(C) {};
                  \draw[sync] (A) to (X1);
                  \draw[sync] (X1) to (X2);
                  \draw[fifo] (X1) to (X3);
                  \draw[sync] (X2) to (B);
             	  \draw[sync] (X3) to (C);
            \end{tikzpicture}
            &
            \raisebox{3pt}{
            \begin{tikzpicture}[->,>=stealth]
                  \tikzstyle{every state}=[draw=black,text=black,minimum size=10pt,node distance=1.6cm]
                  \node[state] (s1) {};
                  \node[state] (s2) [right of=s1] {};
                  \path (s1) edge [bend left] node[above] {$\{A, B\}$} (s2)
                  (s2) edge [bend left] node [below] {$\{C\}$} (s1);
            \end{tikzpicture}}
            \\
            Reo model & Constraint automaton
        \end{tabular}
   }
   \end{tabular}
}
\caption{Specifications of a Reo connector and its wrong implementation (Example 1)}
\label{fig:example1}
\end{figure}
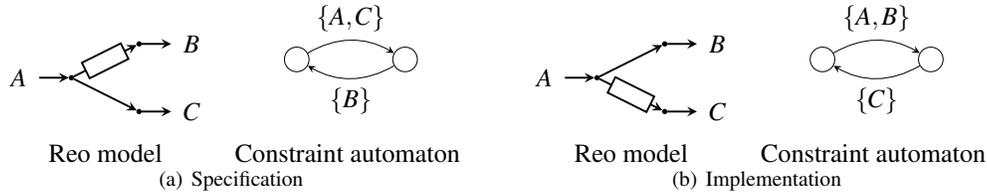

To obtain connector specifications suitable for testing, we combine the idea of explicit representation of pending requests introduced in intentional automata with constraint automata semantics for Reo. Thus, for every boundary Reo port $A$ we introduce two actions $?A$ and $!A$ that represent an external request for this port to accept or dispense a data item and the actual observation of data flow on the circuit node $A$, respectively. Thus, our representation of boundary nodes in {\mCRL} will be as follows:

\begin{tabular}{r@{$\;=\;$}l}
  $\channel{Merger}$ & $?C \cdot (A|!C + B|!C) \cdot \channel{Merger};$ \\
  $\channel{Replicator}$ & $?A \cdot !A|B|C \cdot \channel{Replicator};$\\
  $\channel{Router}$ & $?A \cdot (!A|B + !A|C) \cdot \channel{Router};$\\
\end{tabular}

\begin{figure}
\centering
    \subfigure[Specification]{
       \scalebox{0.7}{
\begin{tikzpicture}[->,>=stealth',scale=2]
    \tikzstyle{state}=[circle, draw]
    \tikzstyle{initstate}=[state,draw=green]
    \definecolor{currentcolor}{rgb}{1,1,1}
    \node at (-80pt, 30pt) [initstate, fill=currentcolor] (state0) {0};
    \node at (-120pt, -60pt) [state, fill=currentcolor] (state1) {1};
    \node at (0pt, 30pt) [state, fill=currentcolor] (state2) {2};
    \node at (-80pt, -30pt) [state, fill=currentcolor] (state3) {3};
    \node at (-120pt, 0pt) [state, fill=currentcolor] (state4) {4};
    \node at (-40pt, 60pt) [state, fill=currentcolor] (state5) {5};
    \node at (-120pt, 60pt) [state, fill=currentcolor] (state6) {6};
    \node at (-40pt, -60pt) [state, fill=currentcolor] (state7) {7};
    \node at (80pt, 30pt) [state, fill=currentcolor] (state8) {8};
    \node at (0pt, -30pt) [state, fill=currentcolor] (state9) {9};
    \node at (40pt, -60pt) [state, fill=currentcolor] (state10) {10};
    \node at (120pt, 0pt) [state, fill=currentcolor] (state11) {11};
    \node at (80pt, -30pt) [state, fill=currentcolor] (state12) {12};
    \node at (120pt, 60pt) [state, fill=currentcolor] (state13) {13};
    \node at (40pt, 60pt) [state, fill=currentcolor] (state14) {14};
    \node at (120pt, -60pt) [state, fill=currentcolor] (state15) {15};

    \path (state0) edge node[fill=currentcolor] {\{?B,?A\}}    (state1);
    \path (state0) edge node[fill=currentcolor] {\{?C,?A\}}    (state2);
    \path (state0) edge node[fill=currentcolor] {\{?C,?B,?A\}} (state3);
    \path (state0) edge node[fill=currentcolor] {\{?B\}}       (state4);
    \path (state0) edge node[fill=currentcolor] {\{?C\}}       (state5);
    \path (state0) edge node[fill=currentcolor] {\{?C,?B\}}    (state6);
    \path (state0) edge node[fill=currentcolor] {\{?A\}}       (state7);
    \path (state1) edge node[fill=currentcolor] {\{?C\}}       (state3);
    \path (state2) edge node[fill=currentcolor] {\{!C,!A\}}    (state8);
    \path (state2) edge node[fill=currentcolor] {\{!C,!A,?B\}} (state9);
    \path (state2) edge node[fill=currentcolor] {\{?B\}}       (state3);
    \path (state3) edge node[fill=currentcolor] {\{!C,!A\}}    (state9);
    \path (state4) edge node[fill=currentcolor] {\{?C,?A\}}    (state3);
    \path (state4) edge node[fill=currentcolor] {\{?C\}}       (state6);
    \path (state4) edge node[fill=currentcolor] {\{?A\}}       (state1);
    \path (state5) edge node[fill=currentcolor] {\{?B,?A\}}    (state3);
    \path (state5) edge node[fill=currentcolor] {\{?B\}}       (state6);
    \path (state5) edge node[fill=currentcolor] {\{?A\}}       (state2);
    \path (state6) edge node[fill=currentcolor] {\{?A\}}       (state3);
    \path (state7) edge node[fill=currentcolor] {\{?B\}}       (state1);
    \path (state7) edge node[fill=currentcolor] {\{?C\}}       (state2);
    \path (state7) edge node[fill=currentcolor] {\{?C,?B\}}    (state3);
    \path (state8) edge node[fill=currentcolor] {\{?B,?A\}}    (state10);
    \path (state8) edge node[fill=currentcolor] {\{?C,?A\}}    (state11);
    \path (state8) edge node[fill=currentcolor] {\{?C,?B,?A\}} (state12);
    \path (state8) edge node[fill=currentcolor] {\{?B\}}       (state9);
    \path (state8) edge node[fill=currentcolor] {\{?C\}}       (state13);
    \path (state8) edge node[fill=currentcolor] {\{?C,?B\}}    (state14);
    \path (state8) edge node[fill=currentcolor] {\{?A\}}       (state15);
    \path (state9) edge node[fill=currentcolor] {\{?C,?A\}}    (state12);
    \path (state9) edge node[fill=currentcolor] {\{?C\}}       (state14);
    \path (state9) edge node[fill=currentcolor] {\{?A\}}       (state10);
    \path (state9) edge node[fill=currentcolor] {\{!B,?A\}}    (state7);
    \path (state9) edge [bend right=15, near start] node[fill=currentcolor] {\{!B,?C,?A\}} (state2);
    \path (state9) edge node[fill=currentcolor] {\{!B\}}       (state0);
    \path (state9) edge node[fill=currentcolor] {\{!B,?C\}}    (state5);
    \path (state10) edge node[fill=currentcolor] {\{?C\}}      (state12);
    \path (state10) edge node[fill=currentcolor] {\{!B\}}      (state7);
    \path (state10) edge node[fill=currentcolor] {\{!B,?C\}}   (state2);
    \path (state11) edge node[fill=currentcolor] {\{?B\}}      (state12);
    \path (state12) edge node[fill=currentcolor] {\{!B\}}      (state2);
    \path (state13) edge node[fill=currentcolor] {\{?B,?A\}}   (state12);
    \path (state13) edge node[fill=currentcolor] {\{?B\}}      (state14);
    \path (state13) edge node[fill=currentcolor] {\{?A\}}      (state11);
    \path (state14) edge node[fill=currentcolor] {\{?A\}}      (state12);
    \path (state14) edge node[fill=currentcolor] {\{!B,?A\}}   (state2);
    \path (state14) edge node[fill=currentcolor] {\{!B\}}      (state5);
    \path (state15) edge node[fill=currentcolor] {\{?B\}}      (state10);
    \path (state15) edge node[fill=currentcolor] {\{?C\}}      (state11);
    \path (state15) edge node[fill=currentcolor] {\{?C,?B\}}   (state12);
\end{tikzpicture}
}
       \label{fig:iocaSpec}
    }
    \subfigure[Implementation]{
       \scalebox{0.7}{
\begin{tikzpicture}[->,>=stealth',scale=2]
\tikzstyle{state}=[circle, draw]
\tikzstyle{initstate}=[state,draw=green]
\definecolor{currentcolor}{rgb}{1,1,1}

 \node at (-80pt, 30pt) [initstate, fill=currentcolor] (state0) {0};
 \node at (0pt, 30pt) [state, fill=currentcolor] (state1) {2};
 \node at (-120pt, 60pt) [state, fill=currentcolor] (state2) {6};
 \node at (-80pt, -30pt) [state, fill=currentcolor] (state3) {3};
 \node at (-40pt, -60pt) [state, fill=currentcolor] (state4) {7};
 \node at (-120pt, 0pt) [state, fill=currentcolor] (state5) {4};
 \node at (-120pt, -60pt) [state, fill=currentcolor] (state6) {1};
 \node at (-40pt, 60pt) [state, fill=currentcolor] (state7) {5};
 \node at (80pt, -30pt) [state, fill=currentcolor] (state8) {12};
 \node at (0pt, -30pt) [state, fill=currentcolor] (state9) {9};
 \node at (120pt, 0pt) [state, fill=currentcolor] (state10) {11};
 \node at (40pt, 60pt) [state, fill=currentcolor] (state11) {14};
 \node at (80pt, 30pt) [state, fill=currentcolor] (state12) {8};
 \node at (120pt, -60pt) [state, fill=currentcolor] (state13) {15};
 \node at (40pt, -60pt) [state, fill=currentcolor] (state14) {10};
 \node at (120pt, 60pt) [state, fill=currentcolor] (state15) {13};

\path (state0) edge node[fill=currentcolor] {\{?B,!A\}} (state1);
\path (state0) edge node[fill=currentcolor] {\{?C,?A\}} (state2);
\path (state0) edge node[fill=currentcolor] {\{?C,?B,?A\}} (state3);
\path (state0) edge node[fill=currentcolor] {\{?B\}} (state4);
\path (state0) edge node[fill=currentcolor] {\{?C\}} (state5);
\path (state0) edge node[fill=currentcolor] {\{?C,?B\}} (state6);
\path (state0) edge node[fill=currentcolor] {\{?A\}} (state7);
\path (state1) edge node[fill=currentcolor] {\{!B,!A\}} (state8);
\path (state1) edge node[fill=currentcolor] {\{!B,!A,?C\}} (state9);
\path (state1) edge node[fill=currentcolor] {\{?C\}} (state3);
\path (state2) edge node[fill=currentcolor] {\{?B\}} (state3);
\path (state3) edge node[fill=currentcolor] {\{!B,!A\}} (state9);
\path (state4) edge node[fill=currentcolor] {\{?C,?A\}} (state3);
\path (state4) edge node[fill=currentcolor] {\{?C\}} (state6);
\path (state4) edge node[fill=currentcolor] {\{?A\}} (state1);
\path (state5) edge node[fill=currentcolor] {\{?B,?A\}} (state3);
\path (state5) edge node[fill=currentcolor] {\{?B\}} (state6);
\path (state5) edge node[fill=currentcolor] {\{?A\}} (state2);
\path (state6) edge node[fill=currentcolor] {\{?A\}} (state3);
\path (state7) edge node[fill=currentcolor] {\{?B\}} (state1);
\path (state7) edge node[fill=currentcolor] {\{?C\}} (state2);
\path (state7) edge node[fill=currentcolor] {\{?C,?B\}} (state3);
\path (state8) edge node[fill=currentcolor] {\{?B,?A\}} (state10);
\path (state8) edge node[fill=currentcolor] {\{?C,?A\}} (state11);
\path (state8) edge node[fill=currentcolor] {\{?C,?B,?A\}} (state12);
\path (state8) edge node[fill=currentcolor] {\{?B\}} (state13);
\path (state8) edge node[fill=currentcolor] {\{?C\}} (state9);
\path (state8) edge node[fill=currentcolor] {\{?C,?B\}} (state14);
\path (state8) edge node[fill=currentcolor] {\{?A\}} (state15);
\path (state9) edge node[fill=currentcolor] {\{!C,?A\}} (state7);
\path (state9) edge [bend right=15, near start] node[fill=currentcolor] {\{!C,?B,?A\}} (state1);
\path (state9) edge node[fill=currentcolor] {\{!C\}} (state0);
\path (state9) edge node[fill=currentcolor] {\{!C,?B\}} (state4);
\path (state9) edge node[fill=currentcolor] {\{?B,?A\}} (state12);
\path (state9) edge node[fill=currentcolor] {\{?B\}} (state14);
\path (state9) edge node[fill=currentcolor] {\{?A\}} (state11);
\path (state10) edge node[fill=currentcolor] {\{?C\}} (state12);
\path (state11) edge node[fill=currentcolor] {\{!C\}} (state7);
\path (state11) edge node[fill=currentcolor] {\{!C,?B\}} (state1);
\path (state11) edge node[fill=currentcolor] {\{?B\}} (state12);
\path (state12) edge node[fill=currentcolor] {\{!C\}} (state1);
\path (state13) edge node[fill=currentcolor] {\{?C,?A\}} (state12);
\path (state13) edge node[fill=currentcolor] {\{?C\}} (state14);
\path (state13) edge node[fill=currentcolor] {\{?A\}} (state10);
\path (state14) edge node[fill=currentcolor] {\{!C,?A\}} (state1);
\path (state14) edge node[fill=currentcolor] {\{!C\}} (state4);
\path (state14) edge node[fill=currentcolor] {\{?A\}} (state12);
\path (state15) edge node[fill=currentcolor] {\{?B\}} (state10);
\path (state15) edge node[fill=currentcolor] {\{?C\}} (state11);
\path (state15) edge node[fill=currentcolor] {\{?C,?B\}} (state12);
\end{tikzpicture}
}
       \label{fig:iocaImpl}
    }
\caption{Example 1: Constraint automaton with inputs and outputs for the specification of a Reo connector and its wrong implementation}
\label{fig:ioca}
\end{figure}
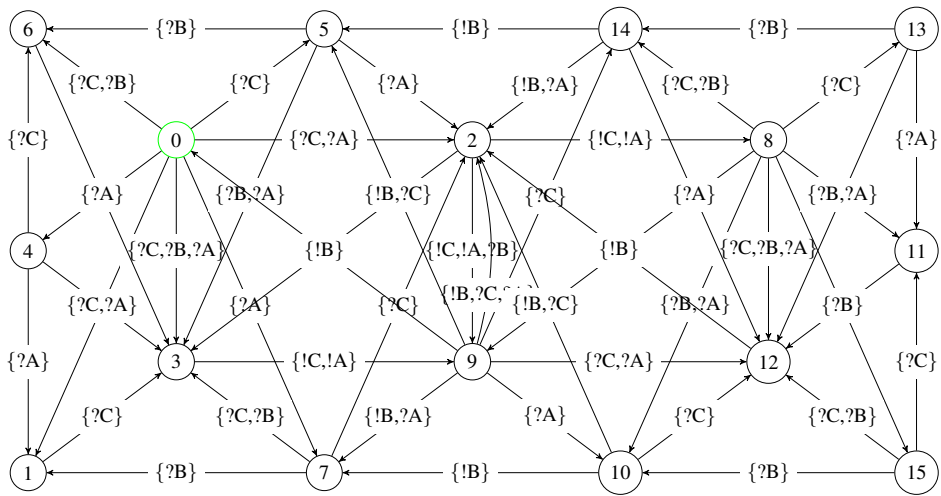
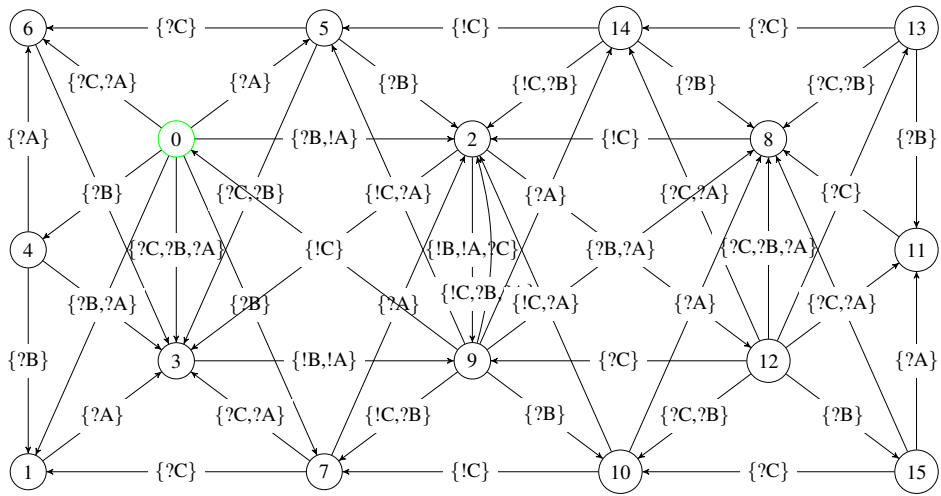

\noindent
Here we assume that the merger node has two internal input ports $A$ and $B$ and a boundary output port $C$ while the replicator and the router nodes have one input boundary port $A$ and two internal output ports $B$ and $C.$ It is not allowed in Reo to have a boundary node which serves both as input and output port. Taking into account that we label input and output events on the same port using different action names (decorated with $?$ and $!$, respectively), we can conclude that for a Reo circuit with all disjoint port names the requirement $L_I \cap L_U = \emptyset$ holds. Figure \ref{fig:ioca} shows constraint automata with inputs and outputs for the specification and implementation of Reo connectors in Example 1.

Aichernig et al.~\cite{AAA+09} developed a tool for testing Reo based on the representation of connectors as designs and specifying them in Maude. The authors claim that testing theories based on finite-state machines are not suitable for testing Reo since in Reo not all input events are followed by output events. While this is true assuming that Reo circuit specifications are provided in the form of basic constraint automata, observe that with our mapping schema we can distinguish a situation when some input item is rejected by a circuit from the case when this item is accepted by the circuit but does not appear on any of the output ports, e.g., destroyed by a {\SyncDrain} or {\LossySync} channels. In fact, any data item supplied by an environment that enters a circuit through an input boundary port $A$ generates an output event $!A.$ Similarly, any output event $!B$ observed on the boundary output port $B$ can only follow the preceding input event $?B$ triggered by the environment. Furthermore, in contrast to earlier approaches based on input/output finite state machines~\cite{BZ99,Pet00}, the \emph{ioco} testing theory allows us to ``observe'' outputs with no data flow on Reo ports (quiescence). We now illustrate why such an extended semantic model is needed to test Reo. In Example 2, the behavior of the circuit in the specification is more general than the behavior of the implemented circuit: for any data input through the input boundary port $A$ in the specification, data flow on the port $B$, port $C$ or both of them simultaneously will be eventually observed. In contrast, in the implementation data flow on ports $B$ and $C$ will be always observed simultaneously. If we generate test cases based on constraint automata, we always observe outputs that are a subset of the admissible outputs in the specification. However, if we explicitly take into account requests from the environment to supply/consume data, we can detect the difference in the circuit implementation. Thus, after observing the input events $?A$ and $?B$ and the output event $!A$, the specification will expect the observation of the action ${!B}$ while the presented wrong implementation will be quiescent.

\begin{figure}
\centering
 \scalebox{0.9}{
    \begin{tabular}{c@{\quad}c@{\quad}c}
     \subfigure[Ignore]{
            \raisebox{12pt}{
            \begin{tikzpicture}[->,>=stealth]
                  \tikzstyle{every state}=[draw=black,text=black,minimum size=10pt,node distance=1.6cm]
                  \node[state] (s1) {};
                  \node[state] (s2) [right of=s1] {};
                  \path (s1) edge node[above] {$\{?A\}$} (s2)
                  (s2) edge[loop above] node[above] {$\{?A\}$} (s2)
                  (s2) edge [bend left] node [below] {$\{!A\}$} (s1);
            \end{tikzpicture}}}
            &
        \subfigure[Overwrite]{
            \begin{tikzpicture}[->,>=stealth]
                  \tikzstyle{every state}=[draw=black,text=black,minimum size=10pt,node distance=1.6cm]
                  \node[state] (s1) {};
                  \node[state] (s2) [right of=s1,label=right:$\quad ... \quad$] {};
                  \node[state] (s3) [right of=s2] {};
                  \node[state] (s4) [right of=s3] {};
                  \path (s1) edge node[above] {$\{?A\}$} (s2)
                  (s3) edge node[above] {$\{?A\}$} (s4)
                  (s2) edge [bend left] node [below, very near start] {$\{!A\}$} (s1)
                  (s3) edge [bend left=45] node [below, very near start] {$\{!A\}$} (s1)                  
                  (s4) edge [bend left=45] node [below] {$\{!A\}$} (s1);
            \end{tikzpicture}}
            &
        \subfigure[Add]{
            \raisebox{12pt}{
            \begin{tikzpicture}[->,>=stealth]
                  \tikzstyle{every state}=[draw=black,text=black,minimum size=10pt,node distance=1.6cm]
                  \node[state] (s1) {};
                  \node[state] (s2) [right of=s1,label=right:$\quad ... \quad$] {};
                  \node[state] (s3) [right of=s2] {};
                  \node[state] (s4) [right of=s3] {};
                  \path (s1) edge node[above] {$\{?A\}$} (s2)
                  (s3) edge node[above] {$\{?A\}$} (s4)
                  (s4) edge[loop above] node[above] {$\{?A\}$} (s4)
                  (s4) edge [bend left] node [below] {$\{!A\}$} (s3)
                  (s2) edge [bend left] node [below] {$\{!A\}$} (s1);
            \end{tikzpicture}}}
   \end{tabular}
 }
\caption{Input request handling}
\label{fig:requestHandling}
\end{figure}
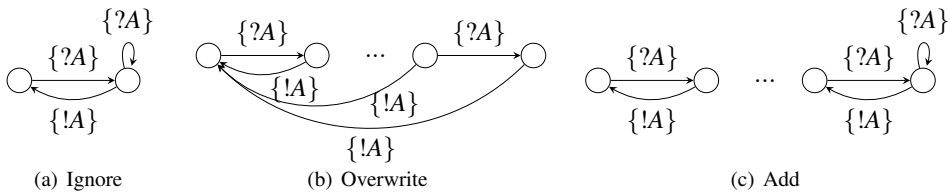

Many existing semantic models for Reo operate at the level of observable data flow on Reo ports and do not specify what happens with possibly multiple requests arriving at  the boundary nodes. There are several strategies to handle these requests: for every port $A$ with a pending request $?A$ on the arrival of another request $?A$ we can (a) ignore the second request, (b) substitute the initial request with the new request, (c) add the second request to the waiting line to be processed by the circuit, e.g., on the FIFO basis. Figure~\ref{fig:requestHandling} shows constraint automata with inputs and outputs-based specifications for the aforementioned strategies. Note that it makes sense to distinguish between the first and the second strategies only for data-aware requests. For data-agnostic circuits it matters only how many requests the circuit needs to process. In the third case, we have to assume that the queue for pending requests is bounded in order to keep the model finite, and after its limit is reached, the further requests are either ignored or overwrite previous ones. What is important is that in all three cases we can see that Reo connector specifications can be represented by constraint automata with inputs and outputs that are input enabled. Based on this observation, we can apply angelic completion for constraint automata with inputs and outputs generated from Reo circuits as discussed above to obtain an input-output constraint automaton without affecting the actual behaviour of the circuit: for any input request $?A$ a subsequent request can influence the behavior of the circuit only after the first request is processed, i.e., action $!A$ is observed, and, thus, adding loops with labels from $L_I$ to each state does not change the semantics of the circuit.

An interesting result follows from the precongruence property for input enabled specifications~\cite{Tre08} (see Proposition 3) and the fact that our generated constraint automata-based specifications are input enabled.

\begin{proposition}
  \label{prop1}
  For any two pairs of connector implementations and specifications, $i_k \in \mathcal{IOCA}(L_{I_k}, L_{U_k})$ and $s_k \in \mathcal{IOCA}(L_{I_k}, L_{U_k}), \, k = 1,2$ with disjoint sets of input/output labels, i.e., $L_{I_1} \cap L_{I_2} = L_{U_1} \cap L_{U_2} = \emptyset,$ it holds that
  $$i_1 \, \mbox{\bf ioco} \, s_1 \mbox{ and } i_2 \, \mbox{\bf ioco} \, s_2 \mbox{ implies } \partial_H (\Gamma_{H \rightarrow \{\tau\}} (i_1 || i_2)) \, \mbox{\bf ioco} \, \partial_H (\Gamma_{H \rightarrow \{\tau\}} (s_1 || s_2)),$$
  where $H = (L_{I_1} \cap L_{I_2}) \cup (L_{U_1} \cap L_{U_2})$ denotes the set of observed actions on their connected ports while $\partial_H (\cdot)$ and $\Gamma_C(\cdot)$ are the {\mCRL} encapsulation and communication operators introduced in Section~\ref{sect:mcrl2}.
\end{proposition}

Practically this means that the product operator on input-output constraint automata preserves the \emph{ioco} relation and testing of Reo connectors can be performed compositionally.

\section{Tool Support}
\label{sect:toolSupport}

To automate testing of Reo, we integrated the JTorX tool into the ECT environment.
JTorX is a Java-based tool to test whether the \emph{ioco} relation holds between a given specification
and a given implementation. JTorX expects the specification to be given in a form of an LTS represented, e.g., in Aldebaran (.aut) or GraphML format.
Thus, we employ our Reo to {\mCRL} conversion framework to generate LTSs that are behaviorally equivalent to constraint automata with inputs and outputs introduced in Section~\ref{sect:testingTheory}. A detailed description of Reo to {\mCRL} mapping plug-in is available in~\cite{KKV11}.
To include input/output actions into an {\mCRL} specification generated from the graphical Reo circuit, select the \emph{I/O actions} check box on the mapping parameters panel. This option can be chosen in combination with coloring and ACA-based mappings. The corresponding {\mCRL} code will appear in the integrated text editor. An LTS with input and output events can be obtained from the generated {\mCRL} code by pressing the \emph{Show LTS} button and saved in the .aut format afterwards. The JTorX tool does not recognize synchronized input and output actions in the form of {\mCRL} multiactions. Therefore, we additionally developed a simple script that converts labels of the form $iA|iB$ and $oA|oB$ into $\{?A, ?B\}$ and $\{!A, !B\},$ respectively. Similarly to {\mCRL}, all actions represented by a set of labels on a single transition in the LTS operated on by JTorX must happen simultaneously, and thus our transformation does not affect the outcome of testing.

The implementation is either given in a form of LTS or it is a real program. In the latter case, JTorX needs to be able to
interact with it, e.g., via the TCP protocol or via an adapter. For testing connector implementations against constraint automata specifications, we can supply both the specification and the implementation in the form of LTS representing their input/output constraint automata semantics. Similarly, for testing implementations of business protocols modeled with Reo, we can obtain LTSs by converting execution code, i.e., BPEL, to Reo~\cite{TVM+08}, and then to {\mCRL}, and, finally, to LTS as described above. However, as this approach requires each translation step to preserve the semantics of the original code, which is not always feasible, a more natural approach would be to develop adapters that execute tests generated by JTorX and observe outputs produced by the real system under test. There is an ongoing work on developing such an adapter for JTorX to communicate with the distributed implementation of Reo in Java~\cite{CPL+10}.

\begin{figure}
\centering
   \includegraphics[width = \textwidth]{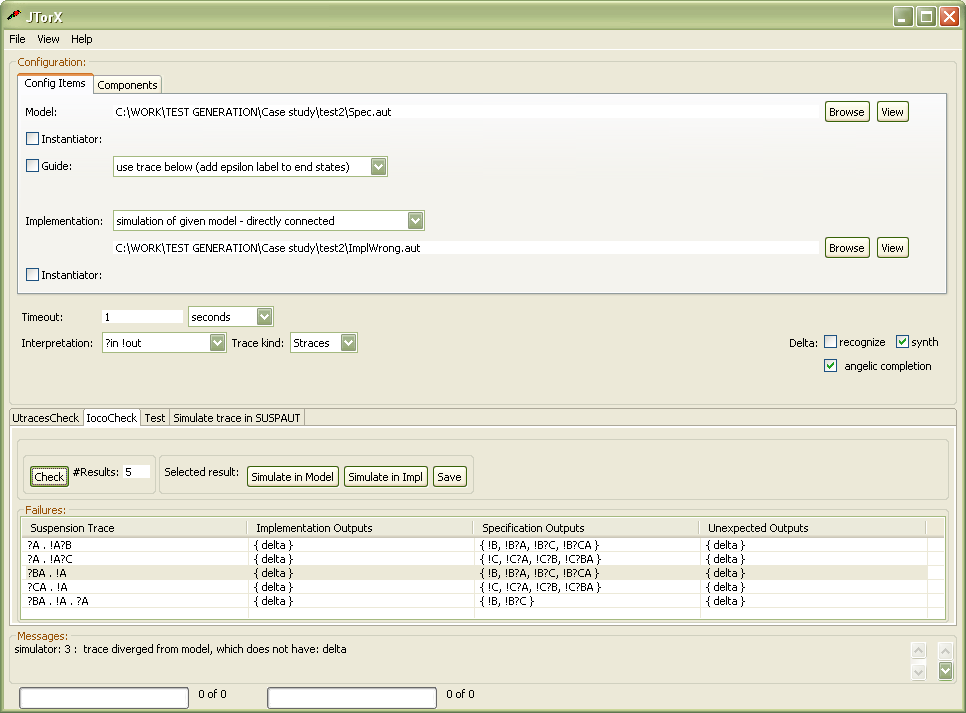}
    \caption{Testing Reo with JTorX: generated test cases for Example 2}
\label{fig:jtorx}
\end{figure}

Figure~\ref{fig:jtorx} shows a screenshot of the JTorX tool with tests generated for Example 2. The highlighted line shows a test case discussed in Section~\ref{sect:testingReo} on which the wrong implementation fails to yield the expected outputs and remains quiescent. Using JTorX, one can simulate test case execution to show traces corresponding to the violated test cases on both specification and implementation LTSs. In our future work, we will develop a plug-in to simulate such test violation traces using Reo animation engine~\cite{ECT}.


\section{Conclusions}
\label{sect:conclusions}

In this paper, we presented an approach to testing models in the Reo coordination language using the \emph{ioco} testing theory. The approach is based on mapping of automata-based semantic models for Reo to the process algebra {\mCRL} and reuse of existing state-space generation and model-based testing tools. We extended the semantic model for Reo with input/output events and showed that the generated specifications are suitable for testing.
In contrast to the previous work on testing Reo~\cite{AAA+09}, where basic connectors are specified equationally and their composition is encoded by means of rewrite rules, no additional effort is required to obtain testable specifications and implementations in our framework. We also expect compositionality of testing Reo with \emph{ioco} to be a useful property that will allow us to assure quality of large process models.

In our future work, we will investigate the applicability of several extensions of \emph{ioco} relation, namely, \emph{symbolic ioco} (\emph{sioco})~\cite{FTW05} and \emph{timed-ioco} (\emph{tioco})~\cite{BB05}, to test time and data-aware Reo circuits.

\bibliographystyle{eptcs}
\bibliography{main}

\end{document}